\documentclass[sigconf,natbib=true]{acmart}

\AtBeginDocument{%
  }

\copyrightyear{2025}
\acmYear{2025}
\setcopyright{cc}
\setcctype{by-nc-sa}
\acmConference[SpatialConnect '25]{The 1st ACM SIGSPATIAL International Workshop on Spatial Intelligence for Smart and Connected Communities}{November 3--6, 2025}{Minneapolis, MN, USA}
\acmBooktitle{The 1st ACM SIGSPATIAL International Workshop on Spatial Intelligence for Smart and Connected Communities (SpatialConnect '25), November 3--6, 2025, Minneapolis, MN, USA}
\acmDOI{10.1145/3764924.3770892}
\acmISBN{979-8-4007-2187-8/2025/11}

\title{Community-Centered Spatial Intelligence for\\Climate Adaptation at Nova Scotia's Eastern Shore}

\author{Gabriel Spadon}
\affiliation{%
  \institution{Faculty of Computer Science\\Dalhousie University}
  \city{Halifax}
  \state{NS}
  \country{Canada}
}
\email{spadon@dal.ca}

\author{Oladapo Oyebode}
\affiliation{%
  \institution{Faculty of Computer Science\\Dalhousie University}
  \city{Halifax}
  \state{NS}
  \country{Canada}
}
\email{oladapo.oyebode@dal.ca}

\author{Camilo M. Botero}
\affiliation{%
  \institution{Faculty of Ind. Engineering\\Dalhousie University}
  \city{Halifax}
  \state{NS}
  \country{Canada}
}
\email{cmbotero@dal.ca}

\author{Tushar Sharma}
\affiliation{%
  \institution{Faculty of Computer Science\\Dalhousie University}
  \city{Halifax}
  \state{NS}
  \country{Canada}
}
\email{tushar@dal.ca}

\author{Floris Goerlandt}
\affiliation{%
  \institution{Faculty of Ind. Engineering\\Dalhousie University}
  \city{Halifax}
  \state{NS}
  \country{Canada}
}
\email{floris.goerlandt@dal.ca}

\author{Ronald Pelot}
\affiliation{%
  \institution{Faculty of Ind. Engineering\\Dalhousie University}
  \city{Halifax}
  \state{NS}
  \country{Canada}
}
\email{ronald.pelot@dal.ca}

\renewcommand{\shortauthors}{Spadon, Oyebode, and Botero et al.}

\setcitestyle{numbers,sort&compress}

\begin{document}

\begin{abstract}
This paper presents an overview of a human-centered initiative aimed at strengthening climate resilience along Nova Scotia's Eastern Shore. This region, a collection of rural villages with deep ties to the sea, faces existential threats from climate change that endanger its way of life. Our project moves beyond a purely technical response, weaving together expertise from Computer Science, Industrial Engineering, and Coastal Geography to co-create tools with the community. By integrating generational knowledge of residents, particularly elders, through the Eastern Shore Citizen Science Coastal Monitoring Network, this project aims to collaborate in building a living digital archive. This effort is hosted under \textit{Dalhousie University} 's \textit{Transforming Climate Action} (TCA) initiative, specifically through its \textit{Transformative Adaptations to Social-Ecological Climate Change Trajectories} (TranSECT) and \textit{TCA Artificial Intelligence} (TCA-AI) projects. This work is driven by a collaboration model in which student teams work directly with residents. We present a detailed project timeline and a replicable model for how technology can support traditional communities, enabling them to navigate climate transformation more effectively.
\end{abstract}



\maketitle

{\small
\begin{table*}[!h]
  \caption{ESCOM's Three Pillars of Data Collection}
  \label{tab:datastreams}
  \begin{tabular}{p{0.2\linewidth} p{0.25\linewidth} p{0.45\linewidth}}
    \toprule
    Data Stream & Methodology & Purpose and Impact\\
    \midrule
    \textbf{Geomorphology}\\\textit{Beach Profiles} & Monthly surveys using a Single-User Beach Profiler (SUBP)~\cite{10.2112/06-0665.1}. & To quantitatively track coastal erosion versus climatic seasonal changes. This provides communities with hard evidence to support planning and adaptation decisions.\\
    \textbf{Ecology}\\\textit{Vegetation Inventory} & Photographic documentation and identification of plant species using PlantNet App\footnote{\url{https://plantnet.org/en/}}. & To create the first comprehensive catalogue of Nova Scotia's Eastern Shore beach flora. This helps identify invasive species and monitor the health of ecosystems on beaches and dunes.\\
    \textbf{Climatology}\\\textit{Weather Monitoring} & Deployment and maintenance of a network of low-cost weather stations and manual rain gauges. & To provide local meteorological data, filling gaps in official records. This enables a detailed analysis of storm impacts, weather variability, and their connection to coastal change.\\
  \bottomrule
\end{tabular}
\end{table*}
}

\section{Introduction}
Long before it was included in official maps, the Eastern Shore of Nova Scotia was a living territory, the ancestral and unceded home of the \textit{Mi'kmaq} people, known as \textit{Sipekne'katik} and \textit{Eskikewa’\-kik}~\cite{sipeknekatik_history, cultural_landscapes}. For millennia, they lived in harmony with the tides, the seasons, and the abundant life of the coast~\cite{pastore1998}. European settlement brought changes, with Canso emerging as a fishing station in the early seventeenth century~\cite{parkscanada_canso}. Over the next three hundred years, a unique culture took root. Small, hardy villages sprang up in sheltered coves and harbors, their existence dictated by the ocean~\cite{historicns_fishfarmfamily}.

Life on the Eastern Shore was defined by a culture of profound self-reliance~\cite{munroe1996times, canso2001portrait}. Communities were built around inshore fishing, boat building, and logging~\cite{island_inklings}. The communal fish shed and the lobster cannery were the economic and social hearts of these villages~\cite{ostrea_clam_factory}. Due to the surrounding environment, people of the shore developed an intimate, almost hereditary knowledge of their environment~\cite{Langendijk_2024}. They could read the weather in the clouds, understand the temperament of the tides, and navigate the treacherous and foggy archipelago of over a thousand wild islands that guard the coast~\cite{wild_islands_nsnt}. Their history is deeply ingrained in the landscape and preserved through folklore, sea shanties, and tales of shipwrecks.

Today, Nova Scotia's Eastern Shore faces an economic shift that has led to the decline of traditional industries, prompting many younger people to seek opportunities elsewhere, leaving behind an aging population~\cite{cooperator_lifelong}. At the same time, climate change presents a more insidious peril~\cite{ns_coastal_flooding}. Rising sea levels, intensifying storms, and coastal erosion are already a reality that threaten homes, wharves, and the very land~\cite{nscc_stormsurge}. The project described in this short paper was created from this context, not as an external intervention, but as a collaborative effort to fortify this historic culture against the coming changes. We aim to bridge the profound, intuitive wisdom of the community with the analytical power of modern science, ensuring the future of the Eastern Shore remains rooted in its past.

The primary contributions of this work are the presentation of: \textbf{(1)} a community-driven, socio-technical model for collecting multi-modal coastal data, spanning geomorphology, ecology, and climatology, using accessible, low-cost methodologies co-designed with residents; \textbf{(2)} the architectural design of a living digital archive built to fuse these data streams, featuring a purpose-built spatial database, a gamified user interface to foster engagement, and a planned conversational system for intuitive data access; and \textbf{(3)} a replicable experiential learning model that bridges academia and community needs by deploying student teams for working towards technical tasks, including digital archaeology of historical records, system backend architecture, and user-centric platform design.

The remainder of this paper is organized as follows: Section 2 introduces the community-driven network. Section 3 describes the Method of Shared Seeing, detailing the data collection process. Section 4 situates the project within its broader institutional framework at Dalhousie University. Section 5 presents the unique experiential learning model that powers the technical development, outlining the roles of student teams. Section 6 provides the phased project timeline, and Section 7 concludes by discussing projected goals.

\section{Collaborative Coastal Monitoring}

At the center of this initiative is the Eastern Shore Citizen Science Coastal Monitoring Network (ESCOM). In this paper, citizen scientists are non-professional volunteers (primarily seniors) who collaborate with researchers to help define questions, collect and analyze data, and interpret results, thereby generating new scientific knowledge~\cite{lawpolicy}. The program emerged not from academic laboratories but from the coastal landscapes it now monitors \cite{deanery_join2025}. Originating in a 2024 pilot led by The Rev. Marian Lucas-Jeferies and Dr. Camilo M. Botero with the intent of connecting science and community \cite{EBERHARDT2022104733}, the network has since matured both physically and digitally \cite{10.2112/JCOASTRES-D-21-00034.1} into a structured and locally embedded effort dedicated to advancing coastal and climate literacy and resilience.

The network's approach is grounded in a philosophy of direct community empowerment~\cite{cooperator_2025}. Rather than positioning residents as passive participants, the project invites them to become active collaborators in the scientific process. Participants are encouraged to define their research questions, apply systematic methods to collect data, and contribute substantively to shared understandings of climate risks. In doing so, the network not only generates valuable observational datasets but also reinforces the principle that local knowledge holders are essential players in climate governance.

The enduring strength of this not-for-profit network lies in its local and regional partnerships. While it receives scientific support from Dalhousie University, its operational capacity is anchored within the social infrastructure of Nova Scotia's Eastern Shore. In the early stages, the Diocesan Environment Network of the Anglican Church played a central role in building trust and fostering engagement, particularly among older residents whose knowledge and experience are key for long-term environmental stewardship. At the same time, the Rural Community Foundation of Nova Scotia provided financial support to reinforce these first steps.

Local institutions are crucial in this ecosystem~\cite{deanery_join2025, diocesan_project2024}. Nonprofit organizations such as The Deanery Project and the Port Bickerton and Area Planning Association, alongside the Anglican Parish of Port Dufferin and the five schools in the area, provide spaces for recruitment, training, and sustained dialogue. These community anchors enable a model of knowledge co-production that deliberately integrates historical insight with contemporary technical skills. Elders offer decades of observational experience and cultural memory, while younger participants contribute digital expertise and novel analytical perspectives. Through this intergenerational collaboration, the network cultivates a living record of coastal change, one that is deeply informed by place, community, and shared responsibility.

\section{Method of Shared Seeing}

ESCOM relies on a simple yet scientifically effective technique for data collection. Its approach democratizes this method, enabling residents to gather rigorous, quantitative data with minimal training and cost, thereby transforming personal observation into a collective scientific endeavor. This process is detailed in Table~\ref{tab:datastreams}.

\subsection{Beach Profile Surveys}

Every month, volunteer teams conduct beach profile surveys. They use a Single-User Beach Profiler (SUBP)~\cite{10.2112/06-0665.1}, constructed from everyday materials such as curtain rods, wing nuts, and a pocket level, to measure the cross-sectional shape of their local beach. This low-tech application of the SUBP method is particularly precise and can detect subtle changes in sand elevation~\cite{10.2112/JCOASTRES-D-23A-00011.1}. This ability enables the community to distinguish between normal seasonal fluctuations and long-term erosion along the coastline. The collected data provides a solid, empirical baseline for assessing the impact of storms or human interventions on nearby coastal communities.

\subsection{Environmental Monitoring}

Volunteers systematically document plant species along the shore, contributing to an inventory of beach flora in the Eastern Shore of Nova Scotia. This work helps track the spread of invasive species while identifying and protecting native plants, such as \textit{Ammophila breviligulata and Honckenya peploides}, that are essential for dune stabilization. Furthermore, ESCOM has established a dense network of over fifteen new low-cost personal weather stations, effectively quadrupling meteorological coverage in some areas. It is complemented by a network of manual rain gauges connected to the Community Collaborative Rain, Hail, and Snow initiative (CoCoRaHS)\footnote{\url{https://www.cocorahs.org/Canada.aspx}}, which records daily pluviosity and snowfall. This citizen weather network allows for climatological analyses at a local level. When a storm hits, the community has its detailed records of rainfall and wind speeds, providing context for any observed erosion or flooding. By running these three streams concurrently (geomorphological, biological, and meteorological), ESCOM creates a detailed picture of coastal change. The use of KoboToolbox\footnote{\url{https://www.kobotoolbox.org/}} for offline data collection has been a key for enabling volunteers in remote areas with unreliable internet to record observations, ensuring no data loss.

\section{Institutional Framework}

The Eastern Shore Citizen Science Coastal Monitoring Network is anchored with Transforming Climate Action (TCA), a \$154 million, nationwide collaboration dedicated to ocean-based climate solutions. Within TCA, the Transformative Adaptations to Social-Ecological Climate-Change Trajectories (TranSECT) stream intends to leverage case studies to explore how coastal communities can adapt to climate risks by uniting environmental science, governance, and community action across multiple disciplines and timescales. Complementing this work, the TCA Artificial Intelligence group works towards developing models, mining community-collected data, and preparing open knowledge bases to enable both local analyses and conversational interfaces for reasoning over the data. Through TCA, community observations are linked to advanced research on climate risk and adaptation, ensuring that local knowledge remains central to resilience planning. Table~\ref{tab:roles} summarizes the core roles and contributions of the groups involved in this initiative.

{\small
\begin{table}[t]
  \caption{Interdisciplinary Roles and Contributions}
  \label{tab:roles}
  \begin{tabular}{p{0.4\linewidth} p{0.5\linewidth}}
    \toprule
    Partner / Group & Role and Contribution\\
    \midrule
    \textbf{ESCOM members} (40+ people in 10+ locations) & The heart of the initiative. Conduct all field data collection, provide local knowledge, validate tools, and champion the activities within their communities.\\
    \textbf{Faculty of Engineering} (Botero, Goerlandt, Pelot) & Provide expertise in coastal geomorphology, risk governance, and coastal engineering. Lead the citizen science methodology and field training.\\
    \textbf{Faculty of Computer Science} (Spadon, Oyebode, Sharma) & Lead the development of the AI tools, the digital platform, and the human-computer interaction design. Oversee the student development teams.\\
    \textbf{Student Teams} & The project engine. Perform data collection, system architecture, front-end development, and UX design, gaining invaluable real-world skills.\\
  \bottomrule
\end{tabular}
\end{table}
}

{\small
\begin{table*}[h]
    \caption{Project Phased Timeline and Key Deliverables}
    \label{tab:timeline}
    \begin{tabular}{p{0.12\linewidth} p{0.07\linewidth} p{0.71\linewidth}}
    \toprule
    Phase & Timeline & Key Milestones and Deliverables\\
    \midrule
    \textbf{1. Pilot} & 2024 & Establishment of first ESCOM sites. Co-design and testing of the methodology. Initial data collection and community feedback sessions. Proof of concept report.\\
    \textbf{2. Expansion} & 2025 & Expansion of ESCOM network to over 20 villages. Development of backend database and AI chatbot prototype. Student teams engaged. Community Feedback Workshops on Digital Tool Design.\\
    \textbf{3. Deployment} & 2026 & Full beta launch of the integrated digital platform and AI chatbot. Widespread digital literacy training for residents. \\
    \textbf{4. Legacy} & 2027+ & Establishment of a long-term community-based risk governance plan supported by ESCOM and IT platforms developed. Publication of project findings and open source toolkits.\\
    \bottomrule
    \end{tabular}
\end{table*}
}

\section{Digital Shore}

Under the mentorship of faculty leads, the CS students involved in the project are tasked with advancing four essential components, each of which contributes to an integrated digital platform:

\begin{enumerate}
\item \textbf{Historical Data Recovery:} Students conduct systematic digital archaeology, extracting open-source related records from province-wide and federal websites about data related to Nova Scotia (in general). This may include, but is not limited to, online: news reports, municipal documents, and community-submitted materials related to flooding, storm events, and shoreline changes. This work ensures that the project's historical database is grounded in both official sources and local memory from the community members.

\item \textbf{Intelligent Data Infrastructure:} Students design and implement the core data architecture that serves as the project's digital platform. This includes a spatially enabled database capable of integrating diverse data sources. The system is designed to facilitate flexible querying and serve as the foundation for advanced analytical tools.

\item \textbf{Conversational Interfaces and Digital Access:} A central focus of the student teams is the development of a conversational system that provides intuitive access to the platform's information resources. This system combines retrieval-based methods with localized knowledge to allow users to query historical climate events or recent shoreline changes through familiar communication tools such as Telegram or web portals. By prioritizing accessibility, the project reduces technological barriers and enables residents with varying levels of digital literacy to engage with it.

\item \textbf{Persuasive Engagement and Participatory Design:} Students lead the design of community-facing features that foster sustained participation. This includes developing a gamified citizen science approach that rewards contributions through recognition systems, progress tracking, and shared accomplishments. Persuasive computing techniques such as social incentives, milestone celebrations, and peer comparisons are embedded to strengthen long-term engagement. These design choices are guided by principles of fairness, inclusivity, and community-driven motivation.

\end{enumerate}

This model creates a mutually beneficial dynamic. Students gain meaningful experience in civic technology, human-centered design, and interdisciplinary collaboration, while the community benefits from a responsive and continuously evolving digital platform. The approach ensures that technical development remains closely aligned with community needs and that the platform serves as both a practical tool and a model for participatory digital innovation.

\section{Project Timeline}

The project follows a multi-year, phased approach to ensure steady progress, community alignment, and long-term sustainability, detailed in Table~\ref{tab:timeline}. The initial phase focused on building trust and piloting core methods. This included establishing monitoring sites, designing field protocols with community members, and testing the citizen science model. Subsequent phases emphasize network expansion, digital platform development, and community engagement. CS students lead the creation of the database, AI-powered conversational tools, and gamified participation features, all of which are refined through community feedback. Later stages focus on platform deployment and integration, supported by digital literacy workshops. The collected data will inform adaptation planning and risk governance strategies in pilot communities. The final phase emphasizes legacy and long-term risk governance. It includes formalizing partnerships with municipalities, sharing project methodologies, and securing sustained leadership moving forward.

\section{Conclusion}

This short paper discussed a community-driven framework for climate adaptation that integrates citizen science, digital tools, and collaborative learning. The project underlines that effective coastal resilience is not defined by technical sophistication alone, but by the capacity of communities to engage with and sustain meaningful participation in environmental monitoring and risk governance.

By combining historical knowledge, systematic field methods, artificial intelligence, and participatory design, this initiative fosters a digital infrastructure that is both scientifically rigorous and community-based. The work underscores that resilience is not only a function of technological innovation but also of social cohesion, shared knowledge, and long-term community stewardship. The model developed here offers a replicable approach for other regions seeking to align digital climate tools with local priorities.

\begin{acks}
This works as partially funded by the OFI Community Climate Adaptation Fund (CCAF) and the TCA project. We are especially grateful to our partners and collaborators in the Eastern Shore communities, with particular appreciation for the members of ESCOM.
\end{acks}

\balance
\bibliographystyle{unsrtnat}
\bibliography{references}

\end{document}